\documentclass[12pt]{article}
\usepackage{amsfonts}
\usepackage{amssymb}
\usepackage{graphicx}
\usepackage{amsmath}
\begin{document}

\title{The Hodge Dual Symmetry of the Green-Schwarz Superstring in $AdS_{5} \otimes S^{5}$  }
\author{ Chuan-Hua Xiong$^{ab}$
\thanks{
E-mail:chxiong@nwu.edu.cn}\\
$^{a}$Institute of Modern Physics, Northwest University,\\
Xi'an, 710069, China \\
$^{b}$Department of Modern Physics \\
University of Science and Technology of China, \\
Hefei, Anhui, 230026, P. R. China}
\date{}
\maketitle

\begin{abstract}
The hidden symmetry and an infinite set non-local conserved currents
of the Green-Schwarz superstring on $AdS_5\otimes S^5$ have been
pointed out by Bena et al. In this paper, we shown that the Hodge
dual between the Maurer-Cartan equation and the equation of motion
 gives the hidden symmetry in the
moduli space of Green-Schwarz  superstring.  Thus by twisty
transforming the vielbeins, we can express the  currents of the
paper$^{\cite{bpr}}$ as the Lax connections by a unique spectral
parameter.
\end{abstract}
\section{Introduction}
The study of the type IIB superstring theory on $AdS_5 \otimes S^5$
give us many new results that are useful for further understanding
the AdS/CFT correspondence$^{\cite{mald}}$. Bena  et al
$^{\cite{bpr}}$ found the hidden symmetry and an infinite set
currents of classically conserved current for the Green-Schwarz
superstring$^{\cite{GS}}$ on $AdS_{5}\otimes S^{5},$ such that it
may be exact solvable. Dolan, Nappi and Witten$^{\cite{DNW}}$\ have
describe the equivalences between this integrable structure and the
Yangian symmetry of the nonlocal currents as Bernard's paper$^{\cite{bernard}%
}$.

For a sigma model on coset space G/H, there exists Hodge dual
symmetry between equations of motion and Maurer-Cartan equations. In
term of the Hodge dual symmetry, one can find the flat connection
$A(\lambda)$ which depend on the Lorentz boost parameter $\lambda =
\exp{\varphi}$.  Thus the flat connection $A(\lambda)$  satisfy
\begin{equation}
\partial_{\mu} U(\lambda) =A_{\mu}(\lambda)U(\lambda)\ .
\end{equation}
 The Green-Schwarz superstring in $AdS_5 \otimes S^5$ is described
by the nonlinear sigma model with WZW term where fields take value
in the coset superspace:$\frac{SU(2,2|4)}{SO(4,1)\otimes
SO(5)}$$^{\cite{MT}}$. However, this model is differ from a simply
sigma model because of the $\kappa $ symmetry in the Green-Schwarz
superstring. It is the $\kappa $ symmetry which guarantee the Hodge
duality of the odd vielbeins between Maurer-Cartan equation and
equation of motion although the cosetspace
$\frac{SU(2,2|4)}{SO(4,1)\otimes SO(5)}$ is not a symmetric space.

In this paper, we investigate the Hodge duality between the
Maurer-Cartan equations  and the equations of motion  and obtain the
Lax-matrix by using the twisted dual transformation which represents
a dressing symmetry for Green-Schwarz string embedding into
$AdS_{5}\otimes S^{5}$.
\section{Hodge dual symmetry of the nonlinear model on symmetric space $G/H$}

Given the group $G$ and defined $H$ to be its stability subgroup, we
can obtain the coset space, $G/H$
\begin{equation}
M = \frac{G}{H}
\end{equation}
and $M$ should be a Riemannian manifold on which $G$ act by
isometries. If $\mathcal{G}$ is the Lie algebra of $G$ and
$\mathcal{H} \subset \mathcal{G}$ denotes the Lie algebra of $H
\subset G$, we have the following direct decomposition:
\begin{equation}
\mathcal{G} = \mathcal{H} \oplus \mathcal{K}
\end{equation}

In particular, $H$ invariance of this decomposition implies
\begin{equation}
[\mathcal{H},\mathcal{H}] \subset \mathcal{H},\quad  [\mathcal{H},
\mathcal{K}] \subset \mathcal{K}.
\end{equation}
If
\begin{equation}
[\mathcal{K},\mathcal{K}] \subset \mathcal{H},
\end{equation}
the coset space $M$ is called symmetric space.

We consider the nonlinear coset model where the field take the
values in the coset $M = G/H$. Define the left-invariant current
\begin{equation}
j = G^{-1} dG=  h + k\ ,
\end{equation}
we also have
\begin{equation}
[ k,k ] = h \ ,\quad [h, h] = h
\end{equation}
and
\begin{equation}
[ h, k ]=k\ .
\end{equation}

The action of this coset model is described by
\begin{equation}
L\propto {\rm Tr}(k_\mu k^{\mu}).
\end{equation}
From the action, we obtain the equation of motion
\begin{equation}
D_{\mu}k^{\mu} = 0
\end{equation}
where
\begin{equation}
D_{\mu} = \partial_{\mu} + h_\mu .
\end{equation}
 In the view of the geometry, there is the inherent structure
 equation, i.e. Maurer-Cartan equation:
\begin{equation}
dj+j\wedge j=0 \ .
\end{equation}
Inducing this equation to the worldsheet, we obtain
\begin{equation}
D_{\mu}k_{\nu}-D_{\nu}k_{\mu}=0 \ .
\end{equation}
It equal to
\begin{equation}\label{m1}
\epsilon^{\mu\nu}D_{\mu}k_{\nu}=0 \ .
\end{equation}
Defined the Hodge dual as
\begin{equation}
*k^{\mu}=\epsilon^{\mu\nu}k_{\nu} \ ,
\end{equation}
then the eq.(\ref{m1}) becomes
 \begin{equation}
D_{\mu }*k^{\mu }=0 \ .
\end{equation}
It is easy to find that the equation of motion  becomes
Maurer-Cartan equation with the $k^{\mu}$ replacing by $*k^{\mu}$.

We can combine the $k$ and the $*k$ with the $\varphi$-depended
parameter owing to the Lorentz invariance:
\begin{equation}
\left(
\begin{array}{c}
\widetilde{k}^{\mu} \\
*\widetilde{k}^{\mu}%
\end{array}%
\right) =\left(
\begin{array}{cc}
\cosh \varphi  & \sinh \varphi  \\
\sinh \varphi  & \cosh \varphi
\end{array}%
\right) \left(
\begin{array}{c}
k^{\mu} \\
*k^{\mu}
\end{array}%
\right)\ .
\end{equation}
It is easy to check that $\widetilde{k}^{\mu}$  and
$*\widetilde{k}^{\mu}$ also satisfy the Maurer-Cartan equation and
the equation of motion respectively.

If we define the projection operator of worldsheet as
\begin{equation}
P_{\pm }^{\mu\nu}= \frac{1}{2}(g^{\mu\nu}\pm
\frac{1}{\sqrt{g}}\epsilon ^{\mu\nu}),
\end{equation}
then the  $\widetilde{K}^{\mu}$  and $*\widetilde{K}^{\mu}$ can be
rewritten as follow:
\begin{eqnarray}
\widetilde{k}^{\mu}& = & \exp\varphi
P_{+}^{\mu\nu}k_{\nu}+\exp(-\varphi)P_{-}^{\mu\nu}k_{\nu}\nonumber\\
& = & \cosh\varphi k_{\nu} +\sinh\varphi *k_{\nu}
\end{eqnarray}
\begin{eqnarray}
*\widetilde{k}^{\mu}&=&\exp\varphi
P_{-}^{\mu\nu}k_{\nu}+\exp(-\varphi)P_{+}^{\mu\nu}k_{\nu}\nonumber\\
& = & \sinh\varphi k_{\nu} +\cosh\varphi *k_{\nu}
\end{eqnarray}
Thus, we can construct the one-parameter families flat connections
$a(\varphi)$
\begin{equation}
a(\varphi)=h+\tilde{k}(\varphi)\ ,
\end{equation}
which satisfy:
\begin{equation}
da+a\wedge a=0 \ .
\end{equation}

Given the flat connections, we have the integrable equation:
\begin{equation}
dU=a(\varphi)U
\end{equation}

\section{The worldsheet Hodge Dual between Maurer-Cartan equation and equation of motion of
Green-Shwarz superstring in $AdS^{5}\otimes S^{5}$}
\subsection{Maurer-Cartan structure equation of coset superspace: $\frac{PSU(2,2|4)}{SO(4,1)\otimes SO(5)}$}
 The $AdS_{5}\otimes S^{5}$ is a coset
space $\frac{SO(4,2)}{SO(4,1)}\otimes \frac{SO(6)}{SO(5)}$. It also
preserves the full supersymmetry of the SUGRA and corresponds to the
maximally supersymmetric background vacuum of IIB SUGRA. Combining
the bosonic $SO(4,2)\otimes SO(6)$ isometry symmetry with the full
supersymmetry, the symmetry turns to be the $PSU(2,2|4)$ acting on
the super coset space $\frac{PSU(2,2|4)}{SO(4,1)\otimes SO(5)}$. In
what follows, we adapt the conventions introduced by$^{\cite{MT}}$.

The left-invariant Cartan 1-forms
\begin{equation}
L^{A}=dX^{M}L_{M}^{A},\quad X^{M}=(x,\theta )
\end{equation}%
are given by
\begin{equation}
G^{-1}dG=L^{A}T_{A}\equiv L^{a}P_{a}+L^{a^{\prime }}P_{a^{\prime }}+\frac{1}{%
2}L^{ab}J_{ab}+\frac{1}{2}L^{a^{\prime }b^{\prime }}J_{a^{\prime }b^{\prime
}}+L^{\alpha \alpha ^{\prime }I}Q_{\alpha \alpha ^{\prime }I}\ ,
\label{cartanform}
\end{equation}%
where $G=G(x,\theta )$ is a coset representative in $PSU(2,2|4)$.
The indices $a, b$ are index of $AdS_5$ and $a^{\prime },b^{\prime
}$ are the indices of $S^5$.

Furthermore, the Cartan 1-forms satisfy the Maurer-Cartan (MC)
equation, i.e. the structure equation of basic one forms on the superspace $%
\frac{PSU(2,2|4)}{SO(4,1)\otimes SO(5)}$
\begin{equation}
d(G^{-1}dG)+(G^{-1}dG)\wedge (G^{-1}dG)=0.
\end{equation}%
Then the super Gauss equations of the induced curvatures $F^{ab}$ and $%
F^{a^{\prime }b^{\prime }}$ defined by $F=dH+H\wedge H$ are
\begin{eqnarray}
F^{ab} &\equiv &dL^{ab}+L^{ac}\wedge L^{cb}=-L^{a}\wedge L^{b}+\epsilon ^{IJ}%
\bar{L}^{I}\gamma ^{ab}\wedge L^{J}, \\
F^{a^{\prime }b^{\prime }} &\equiv &dL^{a^{\prime }b^{\prime }}+L^{a^{\prime
}c^{\prime }}\wedge L^{c^{\prime }b^{\prime }}=L^{a^{\prime }}\wedge
L^{b^{\prime }}-\epsilon ^{IJ}\bar{L}^{I}\gamma ^{a^{\prime }b^{\prime
}}\wedge L^{J}.
\end{eqnarray}%
The super Coddazi equation for the even beins are
\begin{equation}
dL^{a}+L^{b}\wedge L^{ba}=-iL^{I}\gamma ^{a}\wedge L^{I},\quad dL^{a^{\prime
}}+L^{b^{\prime }}\wedge L^{b^{\prime }a^{\prime }}=L^{I}\gamma ^{a^{\prime
}}\wedge L^{I},  \label{Coddazi5}
\end{equation}%
and the super Coddazi equation for the odd beins are
\begin{equation}
dL^{I}-\frac{1}{4}\gamma ^{ab}L^{I}\wedge L^{ab}-\frac{1}{4}\gamma
^{a^{\prime }b^{\prime }}L^{I}\wedge L^{a^{\prime }b^{\prime }}=-\frac{1}{2}%
\gamma ^{a}\epsilon ^{IJ}L^{J}\wedge L^{a}+\frac{1}{2}\epsilon ^{IJ}\gamma
^{a^{\prime }}L^{J}\wedge L^{a^{\prime }}\ .  \label{gc3}
\end{equation}

In the super Gauss equations and the Coddazi equations, the terms on the
left hand side are the usual gauge covariant exterior derivative $d+H\wedge $%
, while the right hand side include the contributions of curvature and
torsion by the fermions.

To embed the IIB superstring into the super coset space
$\mathcal{M}$, we should pull back the Cartan form down to the
world sheet $\Sigma (\sigma ,\tau )$ as
\begin{equation}
L^{A}=L_{M}^{A}dx^{M}=L_{M}^{A}\partial _{i}x^{M}d\sigma
^{i}=L_{i}^{A}d\sigma ^{i}\equiv L_{1}^{A}d\tau +L_{2}^{A}d\sigma \ .
\end{equation}%
Then the Maurer-Cartan 1-form becomes
\begin{equation}
G^{-1}\partial _{i}G=L_{i}^{A}P_{A}=L_{i}^{a}P_{a}+L_{i}^{a^{\prime
}}P_{a^{\prime }}+\frac{1}{2}(L_{i}^{ab}J_{ab}+L_{i}^{a^{\prime }b^{\prime
}}J_{a^{\prime }b^{\prime }})+L_{i}^{\alpha \alpha ^{\prime }I}Q_{\alpha
\alpha ^{\prime }I}\ ,
\end{equation}%
and e.g. the super Coddazi equations for the vector 5-beins (\ref{Coddazi5})
become
\begin{eqnarray}
\epsilon ^{ij}(\partial _{i}L_{j}^{a}+L_{i}^{ab}L_{j}^{b})+i\epsilon ^{ij}%
\bar{L}_i^{I}\gamma ^{a}L_{j}^{I} &=&0\ ,  \label{gc1} \\
\epsilon ^{ij}(\partial _{i}L_{j}^{a^{\prime }}+L_{i}^{a^{\prime
}b^{\prime }}L_{j}^{b^{\prime }})-\epsilon ^{ij}\bar{L}_i^{I}\gamma
^{a^{\prime }}L_{j}^{I} &=&0\ . \label{gc2}
\end{eqnarray}%
The Maurer-Cartan equations for the vielbeins describes the
geometric behavior
for the embedding of the type IIB string world-sheet into the target space $%
AdS_{5}\otimes S^{5}$.

\subsection{The Equation of Motion of Green-Schwarz superstring in $AdS_{5}\otimes S^{5}$}
The $AdS_{5}\otimes S^{5}$ Green-Schwarz superstring action is given
as a nonlinear sigma model on the coset superspace
$\frac{SU(2,2|4)}{SO(4,1)\otimes SO(5)}$  $^{\cite{MT}}$
\begin{equation}
I=-\frac{1}{2}\int_{\partial M_{3}}d^{2}\sigma \sqrt{g}%
g^{ij}(L_{i}^{a}L_{j}^{a}+L_{i}^{a^{\prime }}L_{j}^{a^{\prime
}})+i\int_{M_{3}}s^{IJ}(L^{a}\wedge \bar{L}^{I}\gamma ^{a}\wedge
L^{J}+iL^{a^{\prime }}\wedge \bar{L}^{I}\gamma ^{a^{\prime }}\wedge L^{J}).
\label{action}
\end{equation}%
This action is invariant with respect to the local $\kappa $-transformations
in terms of $\delta x^{a}\equiv \delta X^{M}L_{M}^{a}$, $\delta x^{a^{\prime
}}\equiv \delta X^{M}L_{M}^{a^{\prime }}$, $\delta \theta ^{I}\equiv \delta
X^{M}L_{M}^{I}$
\begin{eqnarray}
&&\delta _{\kappa }x^{a}=0,\quad \delta _{\kappa }x^{a^{\prime }}=0,\quad
\delta _{\kappa }\theta ^{I}=2(L_{i}^{a}\gamma ^{a}-iL_{i}^{a^{\prime
}}\gamma ^{a^{\prime }})\kappa ^{iI} \\
&&\delta _{\kappa }(\sqrt{g}g^{ij})=-16i\sqrt{g}(P_{-}^{jk}\bar{L}%
_{k}^{1}\kappa ^{i1}+P_{+}^{jk}\bar{L}_{k}^{2}\kappa ^{i2})\ .
\end{eqnarray}%
Here $P_{\pm }^{ij}\equiv \frac{1}{2}(g^{ij}\pm \frac{1}{\sqrt{g}}\epsilon
^{ij})$ are the projection operators, and $16$-component spinor $\kappa ^{iI}
$ (the corresponding $32$-component spinor has opposite chirality to that of
$\theta $) satisfy the (anti) self duality constraints
\begin{equation}
P_{-}^{ij}\kappa _{j}^{1}=\kappa ^{i1},\quad P_{+}^{ij}\kappa
_{j}^{2}=\kappa ^{i2},
\end{equation}%
which can be written as $\frac{1}{\sqrt{g}}\epsilon ^{ij}\kappa
_{j}^{1}=-\kappa ^{i1}$, $\frac{1}{\sqrt{g}}\epsilon ^{ij}\kappa
_{j}^{2}=\kappa ^{i2}$, i.e. $\frac{1}{\sqrt{g}}\epsilon ^{ij}\kappa
_{j}^{I}=-S^{IJ}\kappa ^{iJ}$.

From the variation of action (\ref{action}), the equations of motion (EOM)
are obtained$^{\cite{MT}}$
\begin{eqnarray}
\sqrt{g}g^{ij}(\bigtriangledown _{i}L_{j}^{a}+L_{i}^{ab}L_{j}^{b})+i\epsilon
^{ij}s^{IJ}\bar{L}_{i}^{I}\gamma ^{a}L_{j}^{J} &=&0\ ,  \label{EM1} \\
\sqrt{g}g^{ij}(\bigtriangledown _{i}L_{j}^{a^{\prime }}+L_{i}^{a^{\prime
}b^{\prime }}L_{j}^{b^{\prime }})-\epsilon ^{ij}s^{IJ}\bar{L}_{i}^{I}\gamma
^{a^{\prime }}L_{j}^{J} &=&0\ ,  \label{EM2} \\
(\gamma ^{a}L_{i}^{a}+i\gamma ^{a^{\prime }}L_{i}^{a^{\prime }})(\sqrt{g}%
g^{ij}\delta ^{IJ}-\epsilon ^{ij}s^{IJ})L_{j}^{J} &=&0\ ,  \label{EM3}
\end{eqnarray}%
where $\bigtriangledown _{i}$ is the $g_{ij}$-covariant derivative
on the world-sheet $\Sigma (\sigma ,\tau )$.

In term of the relation
\begin{equation}
\bigtriangledown _{i}L^{\hat{a}i}=\frac{1}{\sqrt{g}}\partial _{i}(\sqrt{g}L^{%
\hat{a}i})\ ,
\end{equation}
the above equations of motion (\ref{EM1}) and (\ref{EM2}) can be
rewritten as
\begin{eqnarray}
g^{ij}(\partial _{i}(\sqrt{g}L_{j}^{a})+L_{i}^{ab}L_{j}^{b})+i\epsilon
^{ij}S^{IJ}\bar{L}_{i}^{I}\gamma ^{a}L_{j}^{J} &=&0\ ,  \label{eq1} \\
g^{ij}(\partial _{i}(\sqrt{g}L_{j}^{a^{\prime }})+L_{i}^{a^{\prime
}b^{\prime }}L_{j}^{b^{\prime }})-\epsilon
^{ij}S^{IJ}\bar{L}_{i}^{I}\gamma ^{a^{\prime }}L_{i}^{J} &=&0\ .
\label{eq2}
\end{eqnarray}

Note that
\begin{equation}
\frac{1}{\sqrt{g}}\epsilon^{ij}L_{j}^{1}=-L^{i1}, \quad
\frac{1}{\sqrt{g}}\epsilon^{ij}L_{j}^{2}=L^{i2},
\end{equation}
the equations become
\begin{equation}
\partial_{i}(\sqrt{g}L^{ia})+L_{i}^{ab}\sqrt{g}L^{ib}-i\bar{L}_i^{I}\gamma^{a}
\sqrt{g}L^{iI}= 0 \ ,
\end{equation}
\begin{equation}
\partial_{i}(\sqrt{g}L^{ia^{\prime}})+L_{i}^{a^{\prime}b^{\prime}}\sqrt
{g}L^{ib^{\prime}}+\bar{L}_{i}^{I}\gamma^{a^{\prime}}\sqrt
{g}L^{iI}= 0 \ .
\end{equation}
Here we have used the property of the  $\kappa$ symmetry:
\begin{equation}
P_{-}^{ij}L _{j}^{1}=L ^{i1},\quad P_{+}^{ij}L _{j}^{2}=L ^{i2}.
\end{equation}%
\subsection{The worldsheet Hodge dual transformation between Maurer-Cartan equation(MCE) and equation
of motion(EOM) of superstring on $AdS_5\otimes S^5$}

In order to disclose the duality between the MCE and the EOM, we
first describes the Hodge dual of bosonic and fermionic forms.

As usual, the Hodge dual of the coordinates of world-sheet is given
by
\begin{eqnarray}
\ast(d\sigma^{i}) &=&\frac{-1}{\sqrt{g}}\epsilon ^{ij}dz_{j},\quad
(d\sigma^{1})=d\tau ,\quad (d\sigma^{2})=d\sigma \ ,  \notag \\
\epsilon _{12} &=&-\epsilon _{21}=\epsilon ^{21}=-\epsilon ^{12}=1\ .
\end{eqnarray}%
Thus, the Hodge dual of the even beins $L^{i\hat{a}}$ given by
\begin{equation}
 * L^{i\hat{a}}=-\frac{\epsilon
^{ij}}{\sqrt{g}}L_{j}^{\hat{a}}\
,\quad *L_{i}^{\hat{a}}= {%
\epsilon _{ij}\sqrt{g}}L^{j\hat{a}}\ .
\end{equation}

For the odd beins $L^{iI} (I= 1, 2)$, we have
\begin{equation}
* L^{i1}=\frac{1}{\sqrt{g}}\epsilon^{ij}L_{j}^{1}\ ,
\end{equation}
\begin{equation}
*L^{i2}=\frac{1}{\sqrt{g}}\epsilon^{ij}L_{j}^{2}\ .
\end{equation}

 In term of these duality relations, the MCE (\ref{gc1}) and (\ref{gc2}) can be rewritten as
\begin{equation}
\partial_{i}(\sqrt{g}* L^{ia})+L_i^{ab}(\sqrt{g}* L^{ib})-i\bar
{L}_{i}^{I}\gamma^{a}(\sqrt{g}* L^{iI})=0 \ ,
\end{equation}
\begin{equation}
\partial_{i}(\sqrt{g}\ast L^{i^{a\prime}})+L_{i}^{a^{\prime}b^{\prime}}%
(\sqrt{g}\ast
L^{ib^{\prime}})+\bar{L_{i}}^{I}\gamma^{a^{\prime}}(\sqrt{g}*
L^{iI})=0 \ .
\end{equation}

Applying the Hodge dual transformation:
\begin{equation}
* L^{i\hat{a}}\longleftrightarrow L^{i\hat{a}}\ ,\quad * L^{iI}\longleftrightarrow
L^{iI}\label{hd}
\end{equation}
to these equations, we have
\begin{equation}
\partial_{i}(\sqrt{g}L^{ia})+L_{i}^{ab}\sqrt{g}L^{ib}-i
\bar{L}_{i}^{I}\gamma^{a}(\sqrt{g}L^{iI})=0 \ ,
\end{equation}
\begin{equation}
\partial_{i}(\sqrt{g}L^{ia^{\prime}})+L_{i}^{a^{\prime}b^{\prime}}\sqrt
{g}L^{ib^{\prime}}+\bar{L}_{i}^{I}\gamma^{a^{\prime}}(\sqrt
{g}L^{iI})=0 \ .
\end{equation}
This show that the Maurer-Cartan equation dual to equation of motion
with the transformation(\ref{hd}).

It is clear that the GS string action is invariant under the above dual
transformation. There exists no dual between MC eq.({\ref{gc3}) and EOM eq.(%
\ref{EM3}), because the $L^{I}$ only appears in the Wess-Zumino-Witten term
and has no dynamical contribution to the action. Under the dual
transformation, the 3rd EOM }$\left( \ref{EM3}\right) ${\ changes into
\begin{equation}
(\gamma ^{a}\;^{\ast }L_{i}^{a}+i\gamma ^{a^{\prime }}\;^{\ast
}L_{i}^{a^{\prime }})(\sqrt{g}g^{ij}\delta ^{IJ}-\epsilon
^{ij}s^{IJ})L_{j}^{J}=0\ .
\end{equation}%
Namely, only the first factor takes the dual form. }

{For the $L^{\hat{a}\hat{b}}$, it does not change under duality
since it is not dynamical and does not appear in the Green-Schwarz
string action. }

\section{The twisted dual and integrality}

Now we introduce the twisted dual transformation of vielbeins as
follows. The duality discussed in previous section will be included
as a special case of it. On the world sheet $\Sigma (\sigma ,\tau
)$, it is the re-parametrization transformations along the two
directions of the positive and negative light-cone $\tau \pm \sigma
$ with the scale factors $\lambda =e^{\phi }$ and $\lambda ^{-1}$
correspondently.

For the even vielbein forms $ L^{\hat{a}}$, they will be Lorentz
rotate by $\pm \phi $ oppositely
\begin{equation}
\binom{{\mathcal{L}}^{\hat{a}}}{\ast{\mathcal{L}}^{\hat{a}}}=\left(
\begin{array}
[c]{cc}%
\cosh\varphi & \sinh\varphi\\
\sinh\varphi & \cosh\varphi
\end{array}
\right)  \binom{{{L}}^{\hat{a}}}{\ast{{L}}^{^{\hat{a}}}}\ .
\end{equation}

Thus, we have
\begin{eqnarray}
\mathcal{L}^{i\hat{a}}\left( \lambda \right)
&=& \cosh\varphi{L}^{i\hat{a}}+\sinh\varphi *{L}^{^{i\hat{a}}} \nonumber\\
&=&\frac{1}{2}(\lambda+\lambda^{-1})L^{i\hat{a}}+\frac{1}{2}(\lambda
-\lambda^{-1})*L^{i\hat{a}}\nonumber\\
&=& \lambda
P_{+}^{ij}L_{j}^{\hat{a}}+\lambda^{-1}P_{-}^{ij}L_{j}^{\hat{a}} \ .
\end{eqnarray}

Where we have used $\lambda =\exp {\varphi }$ and  $P_{\pm
}^{ij}\equiv \frac{1}{2}(g^{ij}\pm \frac{1}{\sqrt{g}}\epsilon
^{ij})$ .

The odd vielbein forms $L^{I}$ will rotate oppositely by $\pm
\frac{\varphi}{2} $ together with $\theta ^{J}$\ and $\kappa ^{I}$
\begin{equation}
\binom{{\mathcal{L}}^{I}}{\ast{\mathcal{L}}^{I}}=\left(
\begin{array}
[c]{cc}%
\cosh\frac{\varphi}{2} & \sinh\frac{\varphi}{2}\\
\sinh\frac{\varphi}{2} & \cosh\frac{\varphi}{2}%
\end{array}
\right) \binom{{{L}}^{I}}{\ast {{L}}^{I}},I=1,2.
\end{equation}
 The transformations of odd vielbeins are
\begin{eqnarray}
\mathcal{L}^{iI}\left(\lambda \right)&= &\cosh
\frac{\varphi}{2}L^{iI}+\sinh\frac{\varphi}{2}*{L}^{iI}\nonumber\\
&=&\frac{1}{2}(\lambda ^{\frac{1}{2}}+\lambda ^{-\frac{1}{2}})
L^{iI}+\frac{1}{2}(\lambda^{\frac{1}{2}}-
\lambda^{-\frac{1}{2}})*L^{iI}\nonumber\\
&=&\lambda^{\frac{1}{2}}P_{+}^{ij}L_{j}^{I}+\lambda^{-\frac{1}{2}}P_{-}^{ij}L_{j}^{I}\
.
\end{eqnarray}
 Using the $\kappa$ symmetry
\begin{equation}
P_{-}^{ij}L _{j}^{1}=L ^{i1}, \quad P_{+}^{ij}L _{j}^{2}=L ^{i2}\ ,
\end{equation}
we have
\begin{equation}
\mathcal{L}^{i1}\left(\lambda \right) =
\lambda^{-\frac{1}{2}}L^{i1}\ ,
\end{equation}
\begin{equation}
\mathcal{L}^{i2}\left(\lambda \right) = \lambda^{\frac{1}{2}}L^{i2}.
\end{equation}

It should be point out that the Hodge twisted dual symmetry is not
the symmetry of Metsaev-Tseytlin's action$^{\cite{MT}}$. Actually it
is the hidden symmetry in the moduli space, which is described by
the continuous spectral parameter $\lambda$.

Now we can construct the Lax connection $A_{i}\left( \lambda \right) $ with
the spectral parameter $\lambda $ as
\begin{eqnarray}
A_{i}(\lambda ) &=&H+\mathcal{K}\left( \lambda \right) +\mathcal{F}\left(
\lambda \right)  \notag \\
&=&\frac{1}{2}L_{i}^{\hat{a}\hat{b}}J_{\hat{a}\hat{b}}+\mathcal{L}_{i}^{\hat{%
a}}\left( \lambda \right) P_{\hat{a}}+\mathcal{L}_{i}^{\alpha \alpha
^{\prime }I}\left( \lambda \right) Q_{\alpha \alpha ^{\prime }I}  \notag \\
&=&\frac{1}{2}(L_{i}^{ab}J_{ab}+L_{i}^{a^{\prime }b^{\prime }}J_{a^{\prime
}b^{\prime }})+\frac{1}{2}(\lambda +\lambda
^{-1})(L_{i}^{a}P_{a}+L_{i}^{a^{\prime }}P_{a^{\prime }})  \notag \\
&&+\frac{1}{2}(\lambda -\lambda ^{-1})\Big[\;^{\ast
}(L^{a})_{i}P_{a}+\;^{\ast }(L^{a^{\prime }})_{i}P_{a^{\prime }}\Big]  \notag
\\
&&+\lambda ^{-\frac{1}{2}}L_{i}^{\alpha \alpha ^{\prime }1}Q_{\alpha \alpha
^{\prime }1}+\lambda ^{\frac{1}{2}}L_{i}^{\alpha \alpha ^{\prime
}2}Q_{\alpha \alpha ^{\prime }2}\ ,
\end{eqnarray}%
which looks like the original Cartan form with beins replaced by $\mathcal{L}%
\left( \lambda \right) .$ Such an O(2) transformation, should be
defined in the same covariantly shifted moving frame, (the same
gauge) with covariant constant $N(x)$. Thus the $H$ including in the
covariant derivative will not be twisted. Obviously if $\lambda
=1,\phi =0,$ it is the original Cartan form $\left(
\ref{cartanform}\right) .$ On the "wick rotated" world-sheet we may
take
\begin{equation}
\lambda =\exp {\varphi }=i.
\end{equation}%
Then
\begin{equation}
\mathcal{L}^{i\hat{a}}=i*L^{i\hat{a}},  \label{dt1}
\end{equation}%
Similarly%
\begin{equation}
\mathcal{L}^{i1}=i^{\frac{1}{2}}*L^{i1},\quad \mathcal{L}^{i2}=i^{\frac{1}{%
2}}*L^{i2},  \label{dt2}
\end{equation}%

here i appears, from the difference of sign of Hodge star in
$\mathbb{M}_{2}$ and in $\mathbb{E}_{2}.$ Thus the vierbien
$\mathcal{L}(i)$ becomes simply the Hodge dual of original
vierbien on Euclidean world sheet. And it implies the dual
symmetry of the MCE and the EOM. It is obvious that the Lax
connections $A_{i}(\lambda )$ satisfy the zero curvature (flat
connection) condition:
\begin{equation}
\partial _{i}A_{j}(\lambda )-\partial _{j}A_{i}(\lambda )+[A_{i}(\lambda
),A_{j}(\lambda )]=0,
\end{equation}%
as the linear combination of MCE and EOM i.e. the system is integrable, and
we may introduce the transfer matrices $U\left( \lambda ,\sigma \right) $
\begin{equation}
\partial _{i}U\left( \lambda ,\sigma \right) =A_{i}\left( \lambda ,\sigma
\right) U\left( \lambda ,\sigma \right) .  \label{u}
\end{equation}

\section{Conclusions}
Type IIB Green-Schwarz superstring on $AdS_5 \otimes S^5$ is the
nonlinear Sigma model on the
superspace:$\frac{PSU(2,2|4}{SO(4,1)\otimes SO(5)}=AdS_5 \otimes
S^5\times  fermionic~~ term$. It is well known that there exist the
hidden symmetry and an set infinite conserved current in nonlinear
sigma model. As the papers$^{\cite{h1,h2,h3}}$ point out, the hidden
symmetry and the integrable structure can be obtained by the Hodge
dual transformation.
 In Green-Schwarz superstring on $AdS_5
\otimes S^5$, we find there also exist the dual symmetry between the
Maure-Cartan equation and equation of motion because of the $\kappa
$ symmetry. From the dual symmetry, we obtain the flat connection
with one-parameter and the intergrability of the Green-Schwarz
superstring on $AdS_5 \otimes S^5$.

\section*{Acknowledgments}

We would like to thank Bo-Yu Hou, Bo-Yuan Hou,  Kang-Jie Shi for
helpful discussions. This work  is supported in part by funds from
National Natural Science Foundation of China with grant No.10575080.

\end{document}